\documentclass{aastex62}
\usepackage{amsmath,graphics,graphicx}
\usepackage{easybmat}
\usepackage{arydshln}
\usepackage{bm}
\usepackage{textcomp}
\usepackage{multirow}

\shorttitle{Electron energization in magnetic reconnection outflows  }
\shortauthors{M. Battaglia et al.}

\begin{document}
\title{Electron distribution and energy release in magnetic reconnection outflow regions during the pre-impulsive phase of a solar flare}
\author[0000-0003-1438-9099]{Marina Battaglia}
\affiliation{University of Applied Sciences and Arts Northwestern Switzerland, CH-5210 Windisch, Switzerland }
\author[0000-0002-8078-0902]{Eduard P. Kontar}
\affiliation{School of Physics and Astronomy, SUPA, University of Glasgow, Glasgow G12 8QQ, UK}
\author[0000-0001-7856-084X]{Galina Motorina}
\affiliation{Central Astronomical Observatory at Pulkovo of Russian Academy of Sciences, 196140 Russia}
\affiliation{Ioffe Institute, Polytekhnicheskaya, 26, St. Petersburg, 194021, Russia}
\affiliation{School of Physics and Astronomy, SUPA, University of Glasgow, Glasgow G12 8QQ, UK}

\correspondingauthor{Marina Battaglia}
\email{marina.battaglia@fhnw.ch}

\begin{abstract}
We present observations of electron energization in magnetic reconnection outflows during the pre-impulsive phase of solar flare SOL2012-07-19T05:58. During a time-interval of about 20 minutes, starting 40 minutes before the onset of the impulsive phase, two X-ray sources were observed in the corona, one above the presumed reconnection region and one below. For both of these sources, the mean electron distribution function as a function of time is determined over an energy range from 0.1~keV up to several tens of keV, for the first time. This is done by simultaneous forward fitting of X-ray and EUV data. Imaging spectroscopy with RHESSI provides information on the high-energy tail of the electron distribution in these sources while EUV images from SDO/AIA are used to constrain the low specific electron energies. The measured electron distribution spectrum in the magnetic reconnection outflows
is consistent with a time-evolving kappa-distribution with $\kappa =3.5-5.5$. The spectral evolution suggests that electrons are accelerated to progressively higher energies in the source above the reconnection region, while in the source below, the spectral shape does not change but an overall increase of the emission measure is observed, suggesting density increase due to evaporation. The main mechanisms by which energy is transported away from the source regions are conduction and free-streaming electrons. The latter dominates by more than one order of magnitude and is comparable to typical non-thermal energies during the hard X-ray peak of solar flares, suggesting efficient acceleration even during this early phase of the event.
\end{abstract}


\section{Introduction}\label{intro}
One of the intriguing aspects of solar flares is how efficiently they energize electrons \citep[e.g. reviews by][]{1997JGR...10214631M,2011SSRv..159..107H}. The location of electron acceleration has been suggested in many observations, beginning with the detection of an above-the-looptop hard X-ray source by \citet{Ma94}, to be close to magnetic reconnection outflow regions in the solar corona. A number of  more recent observations support this notion
 \citep[e.g.][]{1996ApJ...459..330F,1999Ap&SS.264..129S,2011LRSP....8....6S,
 Kr13, 2017PhRvL.118o5101K,2018arXiv180805951C}. 
 
 One of the most direct observational signatures of energized electrons is found at X-ray wavelengths. The X-ray bremsstrahlung flux from energetic electrons is proportional to the surrounding density and, as a result, observations of electron energization in the corona are normally difficult to accomplish as the bulk of the hard X-ray (HXR) emission is produced in the dense lower regions of the solar atmosphere and due to the presence of strong soft X-ray (SXR) emission from hot coronal plasma.  Reuven Ramaty High Energy Solar Spectroscopic Imager \citep[RHESSI,][]{Li02} observations using imaging spectroscopy
\citep[e.g.][]{Em03, Ba06, 2013A&A...551A.135S, 2015ApJ...799..129O}
demonstrated the difficulty of observing faint coronal HXR sources in the presence of bright chromospheric footpoint emission. 
Therefore, X-ray studies of electron energization in the corona
are normally done either for dense loops \citep[e.g.][]{Xu08,2011ApJ...730L..22K,2012ApJ...755...32G,2014ApJ...787...86J}, 
or for flares with occulted footpoints \citep[e.g.][]{Kr08,2012ChA&A..36..246B,2017ApJ...835..124E}.
Additional opportunity to study electron energization is provided by the pre-impulsive phase of some events. During this phase, often lasting minutes to several tens of minutes, footpoint HXR emission is typically faint 
or absent while SXR emission increases, indicating heating \citep[e.g.][]{Ac92,Ba09,2014ApJ...789...47B},
even though radio observations of some events suggest the presence of non-thermal particles \citep[e.g.][]{2006PASJ...58L...1A,2012ApJ...758..138A} already at this stage.

While RHESSI observations provide important X-ray spectral diagnostics of the coronal regions, they have a limitation. 
Due to the typically steep power-law shape of the accelerated electron spectrum $\propto E^{-4}$ \citep[e.g.][]{1985SoPh..100..465D},
the total energy is dominated by low-energetic electrons,
whose signatures are  found at photon energies for which the interpretation of the spectrum can be ambiguous and/or to which RHESSI is not sensitive \citep[for a review]{2011SSRv..159..301K}. Indeed, the typical solar corona temperature
is about $0.1$~keV, while RHESSI is sensitive to photons above $\sim 3$~keV
leaving an observational gap at energies between $0.1-3$~keV. One way to overcome this difficulty is to include extreme ultraviolet (EUV) data. \citet{2013ApJ...779..107B} demonstrated how the combination of RHESSI data with observations from the Atmospheric Imaging Assembly (AIA) on the Solar Dynamics Observatory \citep{Le11} allows for inferring the electron distribution over a large energy range from 0.1 keV to several tens of keV. \citet{2015Ge&Ae..55..995M} developed fitting routines to forward fit RHESSI and AIA data simultaneously. With this method, the data from the two instruments are treated as one data set and forward fitted with one model distribution.
As shown by \citet{2015ApJ...815...73B}, this approach leads to a better constraint of the low-energy electrons, resulting in up to a factor of $\sim 30$ lower total energy content of the electron distribution compared with traditional X-ray spectroscopy. Using these combined X-ray-EUV diagnostics, we can characterise  
the electron distribution function in the coronal regions of magnetic reconnection outflow during solar flares. Such observations provide insight into electron energetization in probably turbulent outflows and serve as important constraints for particle acceleration models \citep{2018arXiv180904568D}. 
This is also crucial for understanding the overall flare energetics and for determining magnetic energy release rate, as the bulk of the energy is contained at low energy electrons, below $\sim 20$~keV.
   
Here, we present the first observation of the temporal evolution of electron distributions in magnetic reconnection outflows, focusing on the pre-impulsive phase
 of solar flare SOL2012-07-19T05:58. The pre-impulsive phase of this flare is particularly intriguing since two X-ray sources could be imaged with RHESSI, one above and one below what was identified as the magnetic reconnection region by \citet{2013ApJ...767..168L}. This is very rare and, combined with observations from AIA, allows for studying electron energization in magnetic reconnection outflow regions over an unprecedented range of energies.
In this work, we further investigate the temporal evolution of electron energization and energy transport out of the two sources in the reconnection region, finding that even at this very early stage of the flare, energy transport by free-streaming electrons dominates over conductive energy losses and the overall energy release is considerable. Section \ref{s:analysis} describes the data reduction and summarizes the method that was used for the simultaneous EUV and X-ray analysis.
In section \ref{s:erelease}, the dominant energy transport mechanisms are determined and the energy transport away from the source region by these mechanisms are calculated and compared. The results are discussed in section \ref{s:discuss}.

\section{X-ray and EUV observations of SOL2012-07-19T05:58: magnetic reconnection region }\label{s:analysis}

\begin{figure}[]
\centering
\includegraphics[width=0.5\linewidth]{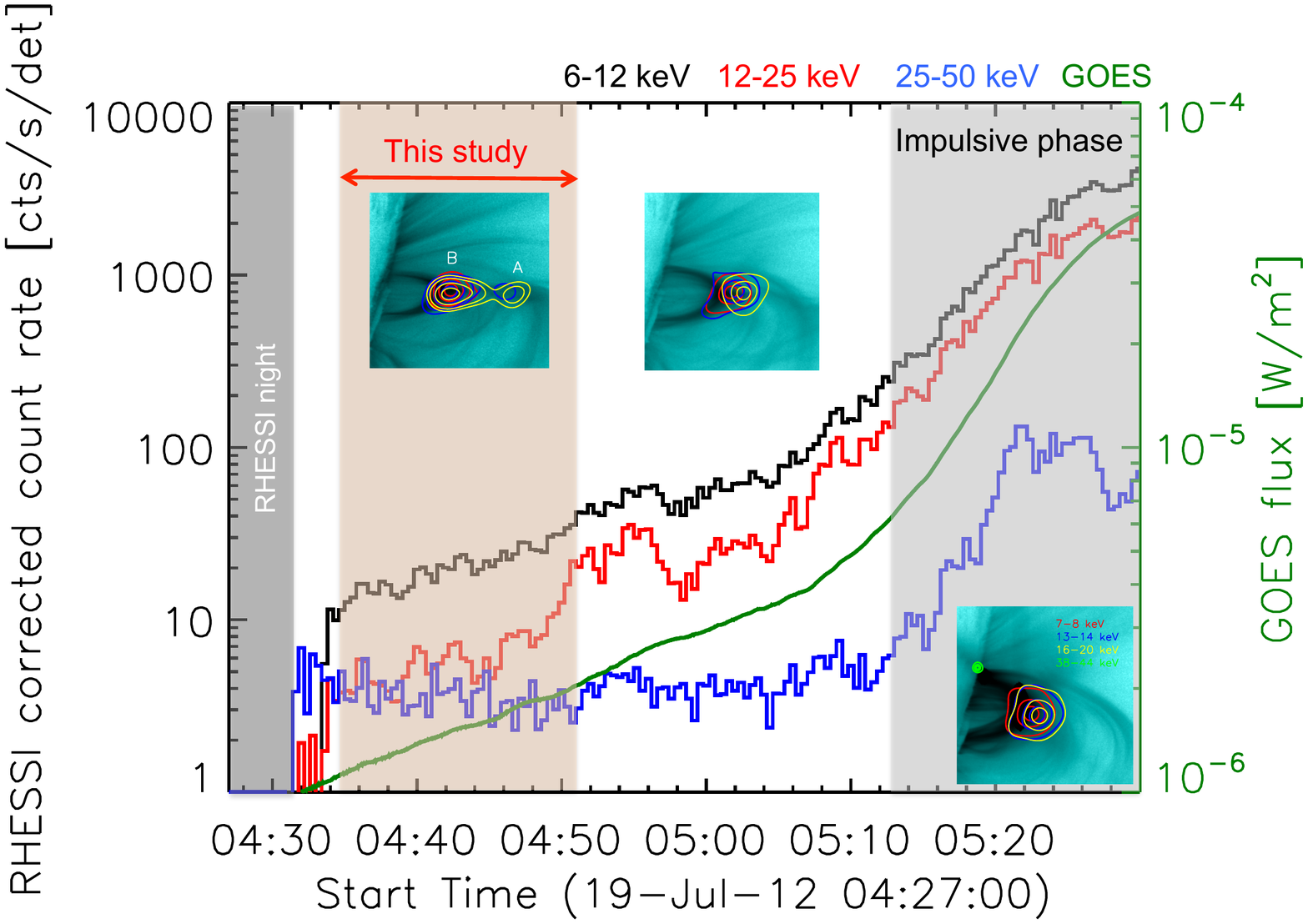}
\caption{RHESSI count-rate lightcurves (corrected for instrumental effects) at 6-12~keV (black), 12-25~keV (red), and 25-50~keV (blue). The green line is the GOES lightcurve. The red arrow indicates the time-range on which this study focuses. Representative images of the flare morphology during three distinct phases are given (see Figure~\ref{fig:sampleimage} for larger images).}
\label{fig:lc}
\end{figure}

The flare SOL2012-07-19T05:58 happened at the west limb and its HXR peak (at 30-60~keV) occurred at $\sim$05:22~UT with the start of the main HXR rise phase at around 05:15~UT. The event displayed a prolonged pre-impulsive phase observed with RHESSI as early as 04:34~UT when RHESSI came out of night. Figure \ref{fig:lc} shows lightcurves of the event in three energy bands of RHESSI and from GOES. 
Several aspects of the event have been studied previously, 
including the temporal and spatial evolution of the whole flare \citep{2013ApJ...767..168L,2014ApJ...786...73S}, 
the properties of the flaring loop \citep{2014Ge&Ae..54..933M} 
and the partition between thermal and non-thermal energies 
during the impulsive phase of the event \citep{2015ApJ...799..129O}, 
particle acceleration during the impulsive phase \citep{2016ApJ...831..119H}, as well as the properties of a hot flux rope structure during the pre-impulsive phase of the flare \citep{2016ApJ...820L..29W} and the CME that was associated with the event \citep{2013ApJ...764..125P}. The focus of the present study is on plasma energization during the pre-impulsive phase of the event near the reconnection region.

\begin{figure*}
\centering
\includegraphics[width=0.98\linewidth]{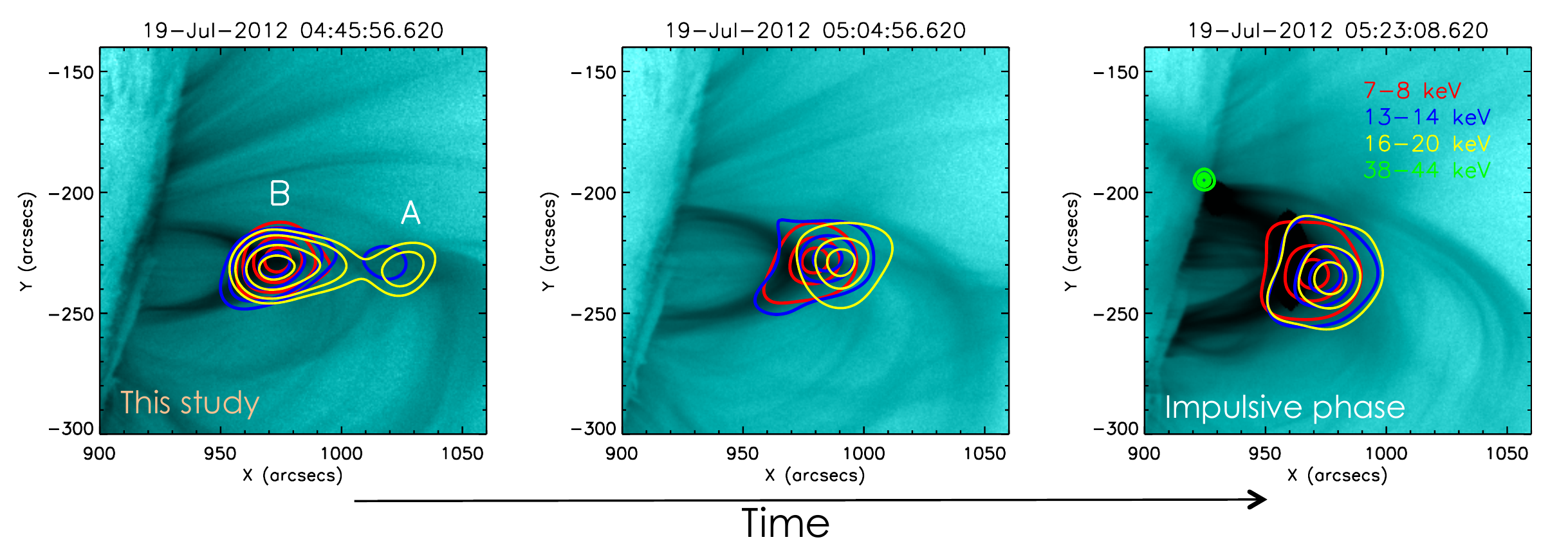}
\caption{AIA 131 \AA\ images at three times (two from before the impulsive phase, one from the impulsive phase). The image on the lefthand side shows a snapshot from the time interval that was analysed in the present study. 40\%, 70\%, 90\% contours from a RHESSI CLEAN image are given in four energy bands: 7-8~keV (red), 13-14~keV (blue), 16-20~keV (yellow), 38-44~keV (green). Two sources, one above the reconnection region (labelled A) and one below (labelled B) were observed during the early pre-impulsive phase until source A disappeared at $\sim$ 04:51~UT.}
\label{fig:sampleimage}
\end{figure*}

Three phases can be distinguished, based on the HXR lightcurves 
and images (Figure \ref{fig:lc}): the first 20 minutes of the pre-impulsive phase when a HXR source was present in the high corona, 
the remaining pre-impulsive phase, and the main flare phase.
Figure \ref{fig:sampleimage} displays AIA images from each of the three 
phases overlaid with contours of X-ray emission at different energies.
In the left image, the two separate coronal X-ray sources are clearly visible, 
one near the top of the EUV loop and one above it.  Following earlier, similar observations 
and the detailed analysis by \citet{2013ApJ...767..168L} we interpret these sources as lying \textit{below} the reconnection region (henceforth referred to as source B) and \textit{above} the reconnection region (henceforth referred to as source A), respectively. In the second image, source A is not visible anymore. 
The third image shows the flare morphology at the onset of the impulsive flare 
phase during which a HXR footpoint was observed in addition to source B. 
In the following we focus on the pre-impulsive phase.

In the next section, we present observations of electron energization 
over a $\sim$ 20 minute interval of pre-impulsive activity, starting 50 minutes before the HXR peak of the event.

\subsection{RHESSI and SDO/AIA data processing}
Using the RHESSI data analysis software\footnote
{\href{https://hesperia.gsfc.nasa.gov/rhessi3/software/software-overview/software-overview/index.html}{https://hesperia.gsfc.nasa.gov/rhessi3/software/software-overview/software-overview/index.html}}
we generated CLEAN images over three minutes integration time 
between 04:34~UT and 04:51~UT, 
with the last image only having an integration time of 2 minutes due to an attenuator state change. The event evolved rather gradually during this phase, therefore the long integration time improves count statistics allowing 
for finer energy binning used in imaging spectroscopy while the temporal evolution can still be followed. 
The spectra of both, source A and source B, were extracted each from within a region that encompasses the whole source. SDO/AIA images at six wavelength bands (94 \AA, 131 \AA, 211 \AA, 335 \AA, 171 \AA, 193 \AA) were averaged over the integration time of the RHESSI images and the data numbers (DN) extracted from the same region as the RHESSI spectrum.

\subsection{Simultaneous fits of RHESSI and AIA data}
To infer the time evolution of the mean electron flux spectrum from both sources we applied the simultaneous fitting method described by \citet{2015Ge&Ae..55..995M,2015ApJ...815...73B}. 
This method uses the fact that any mean electron flux distribution 
$\langle nVF\rangle$ can be described via a differential emission 
measure (DEM) $\xi(T)$:
\begin{equation}
\langle nVF(E)\rangle=\frac{2^{3/2}E}{(\pi m_e)^{1/2}}\int_0^\infty \frac{\xi(T)}{(k_BT)^{3/2}}\exp(-E/(k_BT))dT,\,
\;\;\;\mathrm{[electrons\,keV^{-1}s^{-1}cm^{-2}]}
\end{equation}
where $m_e$ is the electrons mass and $k_B$ the Boltzmann constant. 
The detected signal $g_i$  is given by this DEM multiplied
by the detector response of the instrument 
and the temperature contribution function:
\begin{equation} \label{eq:g_i}
g_i=R_{ij}\xi_j \Delta T_j,
\end{equation}
where $\mathbf{g}=\mathbf{(g^{AIA},g^{RHESSI})}$ are the observed data from AIA in different wavelength channels  and RHESSI at different energies, respectively. 
$R_{ij}$ is the combined temperature and instrument response matrix 
that maps the DEM to the observed data values and  $\Delta T_j$ is the temperature 
bin width. Hence, simultaneous fitting of multi-instrument observations 
is possible by generating a combined temperature response matrix 
and forward fitting it with a model DEM. 
We adopt a DEM model $\xi(T)$ with the following form:
\begin{equation} \label{eq:dem}
\xi (T) =  \frac{EM(\kappa-1.5)^{(\kappa-0.5)}}{\Gamma (\kappa -0.5)T_\kappa} \left(\frac{T_\kappa}{T}\right) ^{\kappa+0.5}
 \times \exp \left(-\frac{T_\kappa}{T}(\kappa-1.5)\right),
\end{equation}
which is equivalent to an electron flux distribution that represents the kappa-distribution \citep{2015ApJ...815...73B}:
\begin{equation}
\langle nVF(E)\rangle =n^2V\frac{2^{3/2}}{\pi (m_e)^{1/2}(k_BT_\kappa)^{1/2}}\frac{\Gamma(\kappa +1)}{(\kappa-1.5)^{1.5}\Gamma(\kappa-0.5)}\frac{E/(k_BT_\kappa)}{(1+E/(k_BT_\kappa)(\kappa-1.5))^{\kappa+1}},
\end{equation}
determined by three parameters:  emission measure $EM$, kinetic temperature $T_\kappa$ and spectral index $\kappa$. This is a natural choice for any source in which one expects electron energization out of an initially Maxwellian distribution, as the $\kappa$-distribution approaches a Maxwellian distribution for $\kappa \rightarrow \infty$ and is dominated by a power-law $<nVF>\propto E^{-\kappa}$ at energies $E>> kT$. Indeed kappa-distributions have been found in a number of coronal flare sources \citep[e.g.][]{2009A&A...497L..13K, 2013ApJ...764....6O,2015ApJ...815...73B}.
We used the DEM from Equation (\ref{eq:dem}) to fit the AIA DN 
and RHESSI count spectrum simultaneously as a function of time in each source.
To evaluate the reliability of the fit and infer uncertainties of the fit parameters, noise, represented by the error of the measured data points multiplied with a random number from a normal distribution with mean zero and standard deviation one, was added to the data and the process repeated twenty times. The standard-deviation 
of the resulting parameters was then taken as uncertainty. 
We note that it is necessary to add one additional fit component to account 
for a low-temperature component in the DEM with a peak at around 1.4~MK. 
It was shown by \citet{2013ApJ...779..107B} 
that this component is predominantly foreground coronal emission, 
also during a flare. Hence we will not include it 
in the further study of the energetics of the reconnection event. 

\begin{figure*}
\centering
\includegraphics[width=0.8 \textwidth]{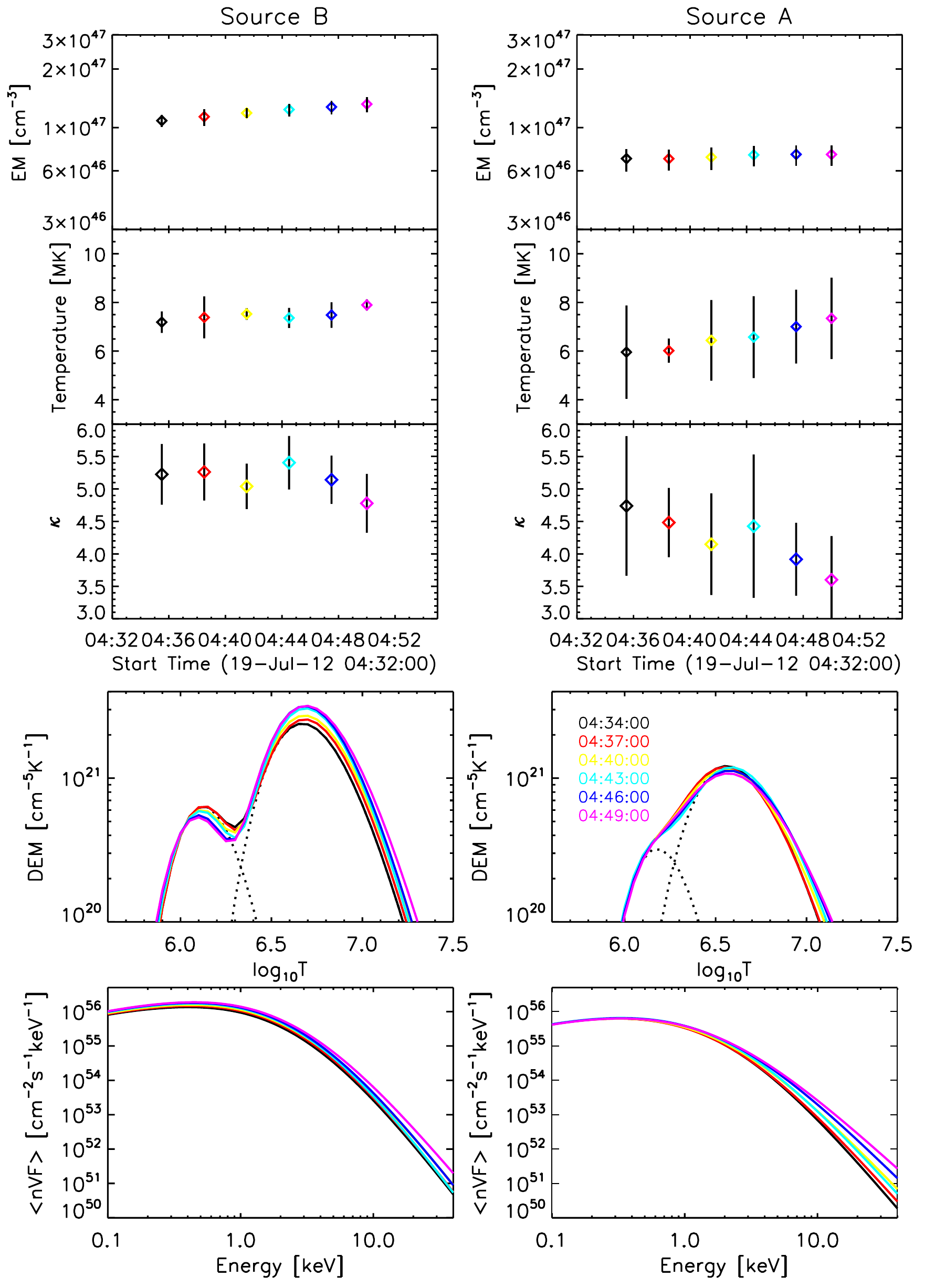}
\caption{Time evolution of fit parameters, differential emission measure and mean electron flux spectrum from simultaneous fits of the $\kappa$-distribution to RHESSI and AIA data. For the source below the reconnetion region (left, source B) and above the reconnection region (right, source A). Top to bottom: emission measure, kinetic temperature $T_\kappa$, $\kappa$-index, DEM, mean electron flux spectrum.}
\label{fig:params}
\end{figure*}

We selected six time intervals and fitted them with the aforementioned
DEM function. The time-evolution of the fit parameters 
of the high-temperature DEM component are shown in Figure \ref{fig:params}.
There is a distinct difference in the evolution of the parameters 
between source A and source B. 
In source A, the $\kappa$ index decreases, 
and the kinetic temperature increases as a function of time,
suggesting more efficient acceleration of electrons to higher energies with time. 
In source B,  $\kappa$ index and temperature remain constant, 
while the emission measure increases, suggesting an increase in density or reduced volume of the emitting plasma, but little change in acceleration or heating. 
The same behaviour can be seen in the DEM. It is also reflected in the resulting volume and density weighted mean electron flux 
spectrum $\langle nVF(E)\rangle$ as shown in Figure \ref{fig:params}. The overall $\langle nVF(E)\rangle$ is larger for source B, but it rises with time across all energies, suggesting energisation of all electrons opposed to source A, where the spectrum changes most notably at higher energies.

\section{Energy release rate and energy transport}\label{s:erelease}
The temporal evolution of both sources suggests continuous energy release. The differences in the two sources, namely a progressively harder spectrum of source A in contrast to the overall increase of emission measure and temperature in source B,  indicate that in source A a considerable amount of the released energy is converted into accelerated particles while in source B heating and increase in density dominates. In either case energy has to be supplied and will successively be transported away 
from the source region. 
To evaluate the energy released during the reconnection, 
we investigate and compare the dominant means of energy transport: 
thermal conduction and through free-streaming non-thermal electrons. 
It should be noted that for the present temperatures and densities, 
radiative losses are several orders of magnitude smaller than either 
of the other two mechanisms and can therefore be neglected.

\subsection{Source geometry and magnetic connectivity}
\begin{figure}
\centering
\includegraphics[width=0.5 \textwidth]{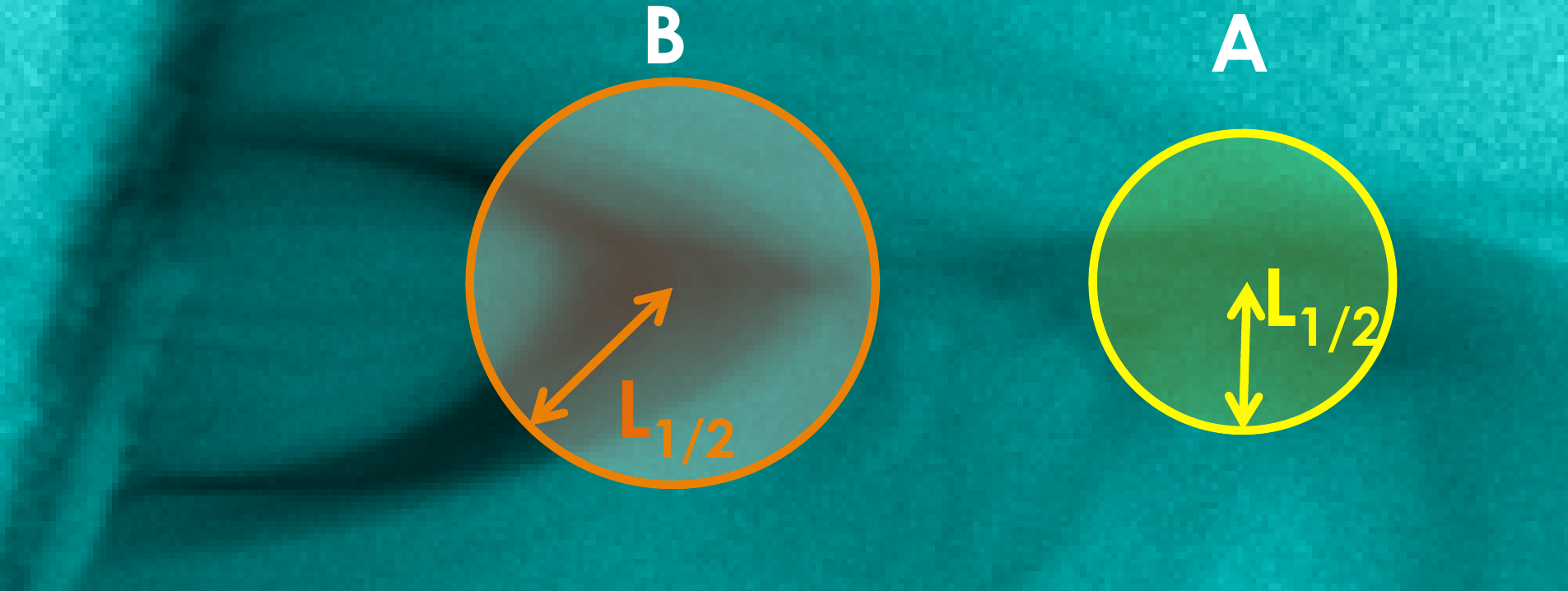}
\caption{Sketch of characteristic length scales used in the energy transport calculations. The background image shows the AIA 131\AA\ EUV emission. The circles indicate the assumed simplified source geometries of sources A and B with their respective radii $L_{1/2}$. }
\label{fig:cartoon}
\end{figure}

For the calculation of both types of energy transport, 
conduction and free-streaming electrons, certain assumptions regarding 
the area through which the energy is lost, the source volume, 
and the distance over which electrons travel have to be made. 
The EUV images suggest a complex geometry with several loop systems, 
while the RHESSI X-ray images show two nearly circular sources, 
depending on photon energy. Assuming that the X-ray sources outline the main electron acceleration regions, we model the sources as spherical in shape 
and estimate their diameters from the X-ray source areas. 
In the case of source A, the diameter is estimated 30 arcsec, 
resulting in a source half-length $L_{1/2}\sim 10^9$ cm 
and a diameter of 40 arcsec results in a half-length 
of $L_{1/2}\sim1.5\times 10^9$ cm for source B .
The volume is calculated from the areas: $V=A^{3/2}$ 
giving $7.2\times 10^{27}$~cm$^3$ for source A
and $1.7\times 10^{28}$~cm$^3$ for source B, respectively.
Figure \ref{fig:cartoon} gives the representation 
of the assumed geometry and indicates the length-scales that 
were estimated from images.

\subsection{Energy transport by conduction and free streaming electrons}
\begin{figure*}
\centering
\includegraphics[width=0.85 \textwidth]{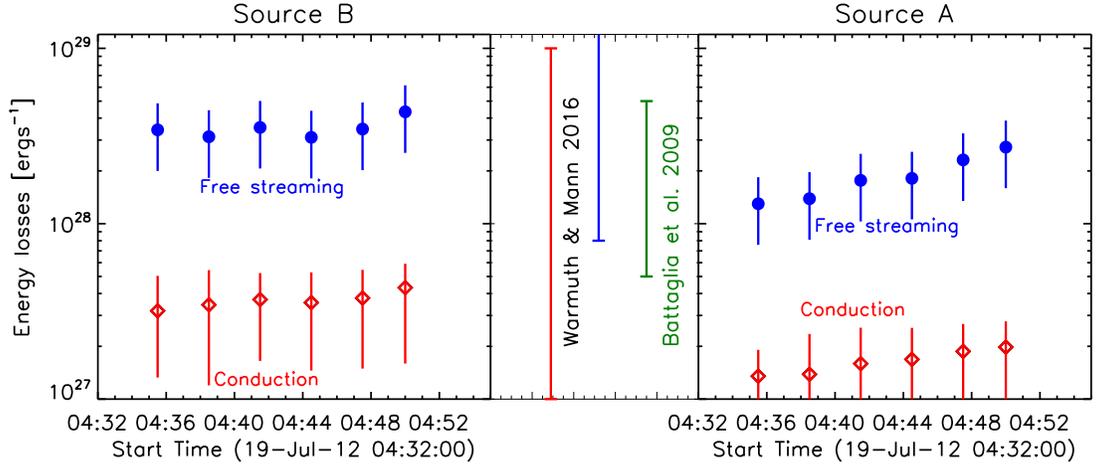}
\caption{Energy losses due to conduction (red) and free streaming electrons (blue) in source B (left) and source A(right). At the center, values from \citet{2016A&A...588A.116W}, where red and blue stands for thermal conduction and non-thermal power, respectively, and \citet{Ba09} (only conduction) are shown. }
\label{fig:theory}
\end{figure*}

The conductive flux in classical Spitzer conductivity \citep{Spbook} 
can be expressed as:
\begin{equation}
L_\text{cond}=10^{-6}A_s\frac{T^{7/2}}{l_{s}}, \quad \mathrm{[erg\,s^{-1}]}
\end{equation}
where $l_s$ is the temperature scale length in cm, T the source temperature in K and $A_s$ the area in cm$^2$ through which the energy is conducted. 
In our case, $l_s$ is assumed to be the source half-length $L_{1/2}$. 
The maximum heat flux a plasma can carry is limited. 
For sufficiently large temperature gradients, such as occur in solar flares, 
the heat flux is expected to reach this limit and saturate \citep{1977PhRvL..39.1270G, Gra80}. 
As shown by \citet{1984PhRvA..30..365C} and \citet{Ba09}, 
the reduced conductive flux can be accounted for by multiplying 
the classical conductive flux with a (temperature dependent) reduction 
factor $\rho = A \exp(-2b(\ln R+c)^2)<1$ with A = 1.01,
b = 0.05,c = 6.63 and $R = \lambda_\text{emf}/l_s$, 
where $\lambda_\text{emf}$ is the electron mean free path. Note that this treatment is only valid as long as $R<1$, i.e. until the free-streaming regime is reached.
In the present case, we do not have a single temperature, 
but a distribution of temperatures given by DEM $\xi (T)$.
Therefore, we calculate the conductive flux weighted 
by the DEM, integrated over all temperatures 
and including the (temperature-dependent) reduction factor $\rho(T)$:
\begin{equation}
L_\text{cond}=10^{-6}\frac{A_s}{L_{1/2}}\frac{1}{EM}\int^{T_\text{emf}}_0T^{7/2}\xi(T)\rho(T)dT.
\end{equation}
The upper integration limit, $T_\text{emf}$ is the temperature 
at which the electron mean free path becomes larger 
than the temperature scale length $\lambda_\text{emf}>L_{1/2}$,
where \citep[e.g.][]{2002ASSL..279.....B}:
\begin{equation}
\lambda_\text{emf}=5.21\times 10^3 \frac{T_\text{emf}^2}{n}\,\;\;\mathrm{[cm]}.
\end{equation}
Hence
\begin{equation}
T_\text{emf}=1.4\times 10^{-2}\sqrt{nL_{1/2}} \,\;\;\; \mathrm{[K]},
\end{equation}
with the standard expression for the density $n=\sqrt{EM/V}$. Electrons with a temperature higher than this threshold can be considered as free-streaming and the energy they carry away calculated from the mean electron flux distribution as:
\begin{equation}
P_\text{free}= \int_{E_\text{emf}}^\infty E F(E)dE=\frac{A_s}{nV}\int_{E_\text{emf}}^\infty E \langle nVF(E)\rangle dE.
\end{equation}
The results from these calculations are summarized in Figure~\ref{fig:theory}. 
The uncertainties are dominated by the source size estimates. 
The error bars given in the figure result from the assumption 
of a 20\% uncertainty on the source half length, volume, and area.
The density in source A increases 
from $2.7\times 10^9$~cm$^{-3}$ 
to $2.9\times 10^9$~cm$^{-3}$
and from $2.8\times 10^9$~cm$^{-3}$  
to $3.4\times 10^9$ ~cm$^{-3}$ 
in source B during the course of the flare.
The resulting threshold temperature for free-streaming electrons rises from 22.9 to 23.4~MK in source A and from 28~MK to 31~MK in source B. 
Expressed in electron energy, this corresponds to electron threshold 
energies of between 2 and  3.5~keV.
\section{Discussion and summary}\label{s:discuss}
We investigated electron energisation and energy loss in the magnetic reconnection outflow regions up to 40 minutes before the impulsive phase of the flare SOL2012-07-19T05:58. As shown in Figure~\ref{fig:theory}, conductive loss rates range between  $1.6\times10^{27}-3.7\times 10^{27}$erg/s. The dominant mechanism of energy transport out of the source region is by free streaming electrons with an average loss rate between $1.8\times10^{28}-3.5\times 10^{28}$erg/s.
To put these numbers into perspective, we compare them with some results from the literature. \cite{2016A&A...588A.116W} analysed the time evolution and energetics of 24 RHESSI flares of GOES classes C to X and found maximum conductive loss rates between $10^{27}-10^{29}$ erg/s. Conductive loss rates specifically during the pre-impulsive phase of flares were reported as $10^9-10^{10}$~erg/cm$^2$/s by \citet{Ba09}. Assuming a source area of $\approx 5\times 10^{18}$ cm$^2$, as done in the present study, this amounts to $5\times 10^{27}-5\times 10^{28}$~erg/s. These values are comparable to the values found in the present study.

A different picture arises when comparing the non-thermal powers.  \cite{2016A&A...588A.116W} found maximum non-thermal powers between $8\times 10^{27}-2\times 10^{29}$~erg/s. Many authors investigated the total energy content, rather than power, hence a comparison is less straightforward. We make an estimate of the non-thermal total energy lost during the 15 minutes observation using the average losses during this period and multiplying them by the duration. This results in a total energy carried by free streaming electrons of $1.7\times 10^{31}$~erg in source A and $3.5\times 10^{31}$ erg in source B. \cite{2016ApJ...832...27A} found total non-thermal energies between $1-200\times 10^{31}$erg analyzing more than 100 GOES M and X-class events. \citet{2012ApJ...759...71E} found similar numbers for 38 M and X-class flares flares with non-thermal energies between $0.4-6\times 10^{31}$erg. These numbers are of the same order as those inferred in the present study, even though they were derived for the flare peak time. This result suggests that even during this early stage of energy release, a lot of energy goes into accelerating electrons. On first glance, this may seem surprising, but it is a direct consequence of the method that was used to infer these energies as it provides good constraints of the low-energy part of the spectrum. As the non-thermal electron spectrum is fairly steep, with typical slopes between 2 to 8, the total power carried by these electrons is determined by the low-energy end of the accelerated population. Our method of simultaneous spectral fitting allows for constraining these energies and, together with the rather low densities in the two analyzed sources, an effective low-energy cutoff as low as 2 keV resulted. This is much lower than the typically used values of around 10 keV and larger. Hence, the values found in these studies have to be considered as lower limits, while the values of the present study give an upper limit. Note that \citet{2016ApJ...832...27A} used an improved method to determine the low-energy cutoff in traditional X-ray spectral fitting, using the warm thick-target bremsstrahlung model by \citet{2015ApJ...809...35K}. 
Hence, their values are better comparable with the numbers found in this study.

It is interesting to compare the inferred energies with the free magnetic energy. One can calculate the (minimum) magnetic field strength assuming 
that the amount of free magnetic energy $ E_{mag}=V{B^2}/{8\pi}$
is converted. Using the average loss rates calculated above and the respective source volumes, one finds $\sim 250$ Gauss in source A and $\sim 230$ Gauss in source B. This number lies within the range (200 to 600 Gauss) given by \cite{2014Ge&Ae..54..933M} right before the impulsive phase of the flare 
presented. 

In summary, we show that considerable electron energization takes place in magnetic reconnection outflows regions up to 40 minutes before the peak of the flare. The observations are broadly consistent with electron acceleration
in likely turbulent reconnection outflows as seen via non-thermal line broadening by \citet{2017PhRvL.118o5101K}.  The spectrum of accelerated electrons in the magnetic reconnection outflows is consistent with a kappa-distribution with a power-law tail spectral index between $-5$ and $-6$ above 2 keV. 
The observations also show efficient heating of the reconnection outflows to temperatures of $6-8$~MK, suggesting that the successful flare acceleration models 
should account for both heating and the formation of power-law tails. 
Both sources show time-evolution at scales longer than the energy loss 
suggesting quasi-stationary energy release of energy. The above the loop-top source shows electron spectrum harderning as the flare progresses. The dominant means of energy loss out of the acceleration region is by free-streaming, low energetic electrons. This not only implies that considerable electron acceleration can take place in flare phases other than the main, impulsive, flare phase but also demonstrates the importance that the pre-impulsive phase plays in overall flare energetics.

\acknowledgements{Acknowledgments}
EPK gratefully acknowledges financial support 
from the STFC Consolidated Grant ST/L000741/1.
GM is supported by the Russian Science Foundation (project no. 16-12-10448). We thank the anonymous reviewer for the comments and suggestions that helped to improve the paper.
\bibliographystyle{apj}
\bibliography{mybib}

\end{document}